\documentclass[aps,prl,twocolumn,groupedaddress,amsmath,amssymb]{revtex4}

\usepackage{graphicx}

\begin{document}

\title{Unidirectional Invisibility induced by {\cal PT}-Symmetric Periodic Structures}

\author{Zin Lin$^1$}
\author{Hamidreza Ramezani$^1$}
\author{Toni Eichelkraut$^2$}
\author{Tsampikos Kottos$^1$}
\author{Hui Cao$^3$}
\author{Demetrios N. Christodoulides$^2$}
\affiliation{$^1$Department of Physics, Wesleyan University, Middletown, Connecticut 06459, USA}
\affiliation{$^2$College of Optics \& Photonics-CREOL, University of Central Florida, Orlando, Florida 32816, USA}
\affiliation{$^3$Department of Applied Physics, Yale University, New Haven, Connecticut 06520, USA}

\date{\today}

\begin{abstract}
We show that parity-time (${\cal PT}$) symmetric Bragg periodic structures, near the spontaneous ${\cal PT}$-symmetry 
breaking point, can act as unidirectional invisible media. In this regime, the reflection from one end is diminished 
while it is enhanced from the other. At the same time the transmission coefficient and phase, are indistinguishable 
from those expected in the absence of a grating. The phenomenon is robust even in the presence of Kerr non-linearities, 
and it can also effectively suppress optical bistabilities.
\end{abstract}


\maketitle

In the last few years considerable research effort has been invested in developing artificial materials-appropriately 
engineered to display properties not found in nature. In the electromagnetic domain, such metamaterials make use of 
their structural composition, which in turn allows them to have complete access of all four quadrants of the real 
$\epsilon-\mu$ plane. Several exotic effects ranging from negative refraction to superlensing and from negative Doppler 
shift to reverse Cherenkov radiation can be envisioned in such systems \cite{PSS06,S07}. Quite recently, the possibility 
of synthesizing a new family of artificial optical materials that instead rely on balanced gain/loss regions has been 
suggested \cite{MGCM08a,MMGC08b,RMGCSK10,GSDMRASC09,RKGC10}. This class of optical structures deliberately exploits 
notions of parity (${\cal P}$)and time (${\cal T}$) symmetry \cite{BB98,BBDJ02,B07} as a means to attain altogether 
new functionalities and optical characteristics \cite{MGCM08a}. Under ${\cal PT}$ symmetry, the creation and absorption 
of photons occurs in a controlled manner, so that the net loss or gain is zero. In optics, ${\cal PT}$ symmetry demands 
that the complex refractive index obeys the condition $n({\vec r})=n^*(-{\vec r})$, in other words the real part of the refractive 
index should be an even function of position, whereas the imaginary part must be odd. ${\cal PT}$-synthetic materials 
can exhibit several intriguing features. These include among others, power oscillations \cite{MGCM08a,RMGCSK10,ZCFK10},
absorption enhanced transmission \cite{GSDMRASC09}, double refraction and non-reciprocity of light propagation 
\cite{MGCM08a}. In the nonlinear domain, such pseudo-Hermitian nonreciprocal effects can be used to realize 
a new generation of on-chip isolators and circulators \cite{RKGC10}. Other exciting results within the
framework of ${\cal PT}$-optics include the study of Bloch oscillations \cite{L09a}, and the realization of coherent
perfect laser absorbers \cite{L10b} and nonlinear switching structures \cite{SXK10}.

To date, most of the studies on optical realizations of ${\cal PT}$ synthetic media have relied on the paraxial approximation 
which maps the scalar wave equation to the Schr\"odinger equation, with the axial wavevector playing the role of energy. 
This formal analogy, allows one to investigate experimentally fundamental ${\cal PT}$ -concepts that may impact several other 
areas, ranging from quantum field theory and mathematical physics \cite{BB98,BBDJ02,B07}, to solid state \cite{BFKS09} 
and atomic physics \cite{GKN08}. Among the various themes that have fascinated researchers, is the existence of spontaneous 
${\cal PT}$ symmetry breaking points (exceptional points) where the eigenvalues of the effective non-Hermitian Hamiltonian 
describing the dynamics of these systems abruptly turn from real to complex \cite{B07}. Recently interest in 
${\cal PT}$ -scattering 
configurations \cite{CDV07,M09,L10d,S10} has been revived in connection with using such devices under a dual role, that of a 
lasing and a perfect coherent absorbing cavity \cite{L10b,CGCS10}.


\begin{figure}[h]
\includegraphics[width=1\columnwidth,keepaspectratio,clip]{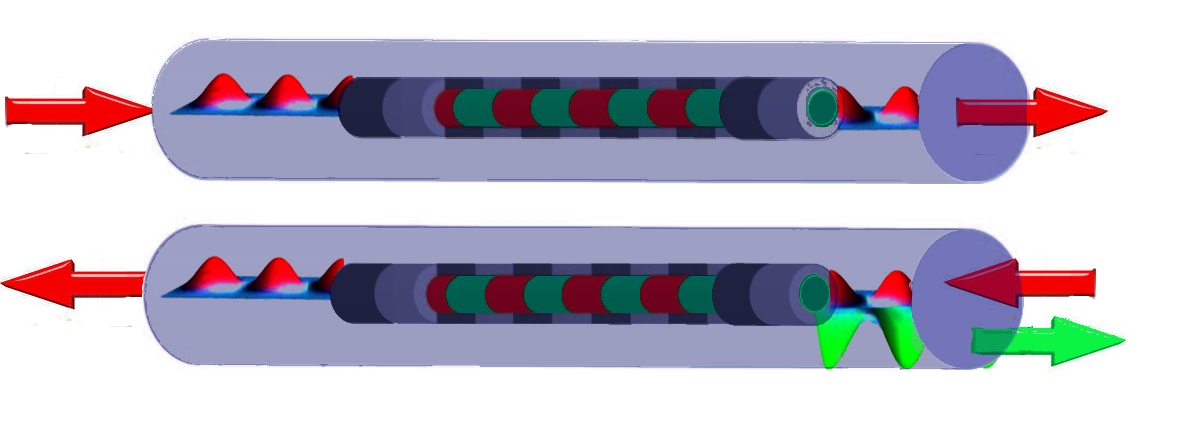}
\caption{
   (color online) Unidirectional invisibility of a ${\cal PT}$-symmetric Bragg scatterer. The wave entering 
from the left (upper figure) does not recognize the existence of the periodic structure and goes through the 
sample entirely unaffected. On the other hand, a wave entering this same grating from the right (lower figure), 
experiences enhanced reflection.
   }
\label{fig1}
\end{figure}

In this Letter we explore the possibility of synthesizing ${\cal PT}$-symmetric objects which can become 
unidirectionally invisible at the exceptional points. In recent years the subject of cloaking physics has 
attracted considerable interest, specifically in connection to transformation optics \cite{PSS06,L06}. Here, 
our notion of invisibility stems from a fundamentally different process. As opposed to surrounding a scatterer 
with a cloak medium, in our case the invisibility arises because of spontaneous ${\cal PT}$-symmetry breaking. 
This is accomplished via a judicious design that involves a 
combination of optical gain and loss regions and the process of index modulation. Specifically, we consider 
scattering from ${\cal PT}$-synthetic Bragg structures (see Fig. \ref{fig1}) and investigate the consequences 
of ${\cal PT}$ symmetry in the scattering process. It is well known that passive gratings (involving no gain 
or loss) can act as high efficiency reflectors around the Bragg wavelength. Instead, we find that at the 
${\cal PT}$ symmetric breaking point, the system is reflectionless over all frequencies around the Bragg 
resonance when light is incident from one side of the structure while from the other side its reflectivity 
is enhanced. Furthermore, we show that in this same regime the transmission phase vanishes-a necessary 
condition for evading detectability. Even more surprisingly, is the fact that these effects persist even 
in the presence of Kerr non-linearities. 

To demonstrate these effects we consider an optical periodic structure or grating 
having a ${\cal PT}$-symmetric refractive index  distribution $n(z)=n_{0} +n_{1} \cos (2 \beta z)+ i n_{2} \sin 
(2 \beta z)$ for $|z|<L/2$. This grating is embedded in a homogeneous medium having a uniform refractive index 
$n_0$ for $|z|>L/2$ (see Fig. \ref{fig1}). Here $n_1$ represents the peak real index contrast and $n_2$ the 
gain/loss periodic distribution. In 
practice, these amplitudes are small, e.g. $n_1,n_2\ll n_0$. The grating wavenumber $\beta$ is related 
to its spatial periodicity $\Lambda$ via $\beta=\pi/\Lambda$ and in the absence of any gain modulation ($n_2=0$) 
the periodic index modulation leads to a Bragg reflection close to the Bragg angular frequency $\omega_{\beta}=c 
\beta /n_{0}$ (where $c$ is the speed of light in vacuum). In this arrangement, a time-harmonic electric field of 
frequency $\omega$ obeys the Helmholtz equation:
\begin{equation}
\label{Helmholtz}
{\partial^2 E (z)\over \partial z^2} + {\omega^2 \over c^2} n^2(z) E(z) = 0\,\,\,.
\end{equation}
For $|z|\geq L/2$, Eq. (\ref{Helmholtz}) admits the solution $E_0^{-}(z)=E_{f}^- \exp(ikz) +
E_{b}^- \exp(-ikz)$ for $z< -L/2$ and $E_0^{+}(z)=E_{f}^+ \exp(ikz) + E_{b}^+ \exp(-ikz)$ for $z> L/2$ where the 
wavevector $k= n_0\omega/c$. The amplitudes of the forward and backward propagating waves outside of the grating domain 
are related through the transfer matrix $M$:
\begin{equation}
\label{transfer}
\left(\begin{array}{c}
E_{f}^+\\
E_{b}^+
\end{array}\right)=
\left(\begin{array}{cc}
M_{11}&M_{12}\\
M_{21}&M_{22}
\end{array}\right)
\left(\begin{array}{c}
E_{f}^-\\
E_{b}^-
\end{array}\right)
\end{equation}
The transmission and reflection amplitudes for left (L) and right (R) incidence waves, can be obtained from the 
boundary
conditions $E_b^+=0$ ($E_f^-=0$) respectively, and are defined as $t_L\equiv {E_f^{+}\over E_f^{-}}$, $r_L\equiv 
{E_b^{-}\over E_f^{-}}$; ($t_R\equiv {E_b^{-}\over E_b^{+}}$; $r_R\equiv {E_f^{+}\over E_b^{+}}$). These can be
expressed in terms of the transfer matrix elements as follows\cite{CDV07,M09}
\begin{equation}
\label{trcoefficients}
t_L=t_R=t={1\over M_{22}}\,\,;\,\, r_L= -{M_{21}\over M_{22}}\,\,; \,\, r_R = {M_{12}\over M_{22}}
\end{equation}
While the transmission for left or right incidence is the same, this is not necessarily the case for the
reflection. From the above relations one can deduce the form of the scattering matrix $S$ \cite{M09} in 
terms of the $M$-matrix elements. For ${\cal PT}$-symmetric systems, the eigenvalues of the $S$-matrix 
either form pairs with reciprocal moduli or they are all unimodular. In the latter case the system is in 
the exact ${\cal PT}$-phase while in the former one it is in the broken-symmetry phase \cite{CDV07,L10b}. 
For the complex periodic structure considered here, the transition from one phase 
to another (spontaneous ${\cal PT}$-symmetry breaking point) takes place when $n_1=n_2$ \cite{RLKCEC11}.

\begin{figure}[h]
\includegraphics[width=1\columnwidth,keepaspectratio,clip]{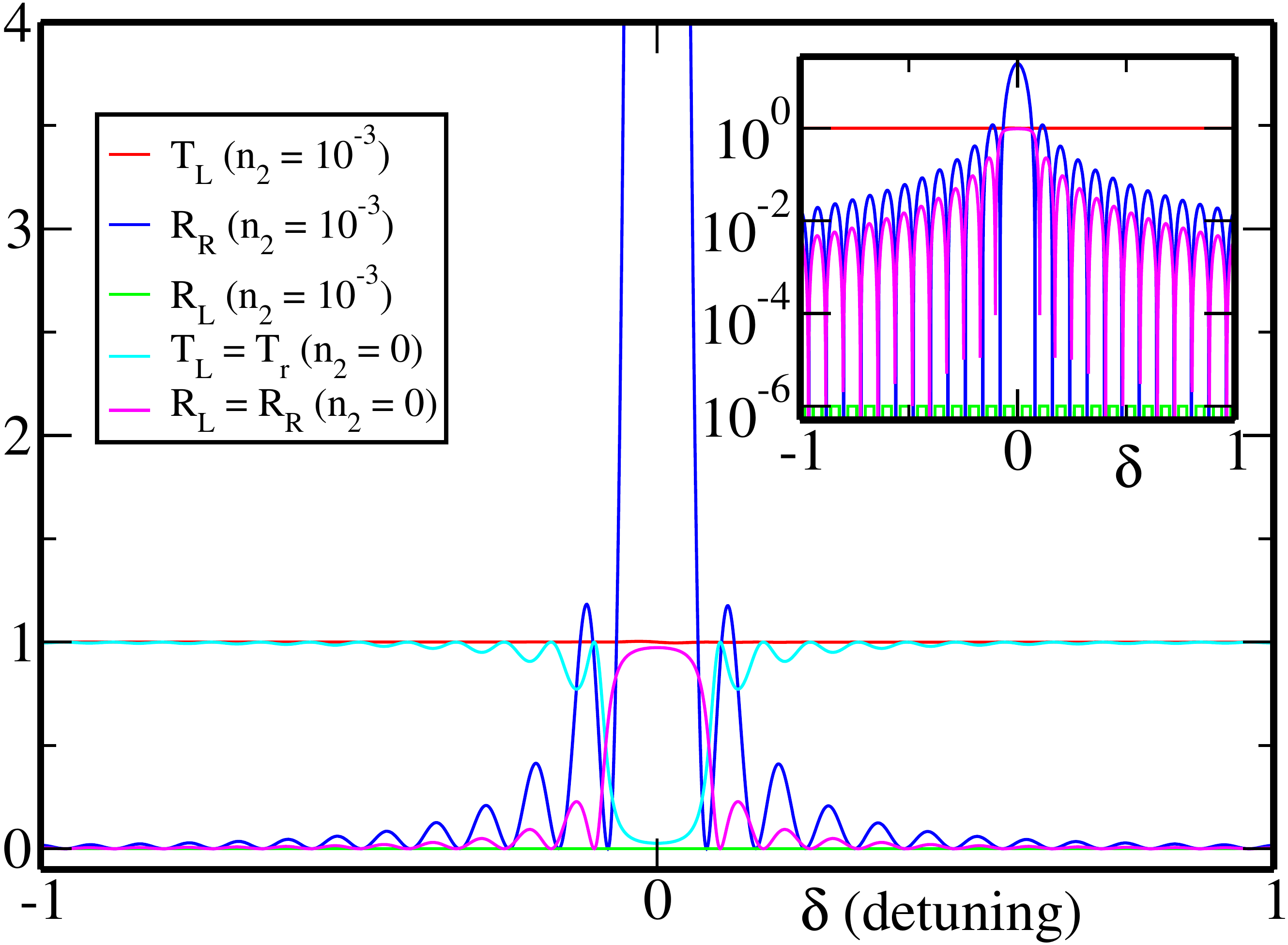}
 \caption{
   \label{fig2}
(color online) Exact numerical evaluation (from Eq. (\ref{Helmholtz})) of transmission $T\equiv |t|^2$ and reflection 
$R=|r|^2$ coefficients for a Bragg grating \cite{numerics}. 
We have used $n_0=1$, $n_1=10^{-3}$, $L=12.5\pi$ and $\beta=100$. In case of a ${\cal PT}$ grating, the system is at the 
exceptional point when $n_2=n_1$. In this case, $R_L$ is diminished (up to $n_{1,2}^2\sim 10^{-6}$ --see inset) 
for a broad frequency band, while $R_R$ is enhanced. These exact numerical results are in excellent agreement with Eqs. 
(\ref{tcoef},\ref{Rright}). 
}
\end{figure}

To analyze this structure we decompose the electric field inside the scattering 
domain $E(z)$, in terms of forward $E_f(z)$ and backward $E_b(z)$ traveling envelopes as 
\begin{equation}
\label{einside}
E(z)=E_f(z) \exp(ikz) + E_b(z) \exp(-ikz).
\end{equation}
Next by employing slowly varying envelopes for the field i.e. $E_f(z)= {\cal E}_f(z) \exp(i\delta z)$
and $E_b(z)= {\cal E}_b(z) \exp(-i\delta z)$, where $\delta=\beta-k$ is the detuning. Substituting these expressions in 
Eq. (\ref{Helmholtz}), and keeping only synchronous terms while eliminating second order corrections in $n_1$,and $n_2$, 
we can then express the field at a point $z$ inside the sample in terms of the field at $z=-L/2$. For $k\approx \beta$ close 
to the Bragg point, we get 
\begin{equation}
\label{cmt}
\left(\begin{array}{c}
E_{f}(z)\\
E_{b}(z)
\end{array}\right)=
e^{iz\delta {\hat \sigma_3}}
{\hat U}
e^{iL\delta {\hat \sigma_3}/2}
\left(\begin{array}{c}
E_{f}(-{L\over 2})\\
E_{b}(-{L\over 2})
\end{array}\right)
\end{equation}
where ${\hat U}= \cos[\lambda (z+L/2)] {\hat 1} - i \sin[\lambda (z+L/2)] {\hat \sigma}\cdot{\hat e}$, 
${\hat \sigma}$ are the Pauli matrices, and 
the unit vector ${\hat e}$ is defined as ${\hat e}=(1/\lambda) (-kn_2/2n_0;-ikn_1/2n_0;\delta)$, while $\lambda=\sqrt{
\delta^2-k^2 (n_1^2-n_2^2)/4n_0^2}$. By imposing continuity of the field at $z=\pm L/2$, Eq.~(\ref{cmt}) 
becomes equivalent to Eq.~(\ref{transfer}). The transmission $T\equiv |t|^2$ and reflection coefficients $R_L\equiv 
|r_l|^2$ and $R_R\equiv |r_r|^2$ are in this case 
\begin{eqnarray}
\label{tcoef}
T = {|\lambda|^2 \over |\lambda|^2 \cos^2(\lambda L) + \delta^2|\sin(\lambda L)|^2} \quad\quad\quad\quad\quad\\
R_L = {(n_1-n_2)^2k^2/4n_0^2\over \delta^2 + |\lambda\cot(\lambda L)|^2} \quad;\quad
R_R = {(n_1+n_2)^2k^2/4n_0^2\over \delta^2 + |\lambda\cot(\lambda L)|^2}\nonumber
\end{eqnarray}
For $n_2=0$ one recovers the standard scattering features of periodic Bragg structures. Namely, $R_L=R_R$, while 
close to the Bragg point $\delta=0$ the reflection/transmission becomes unity/zero (in the large $L$-limit), see 
Fig. \ref{fig2}. Instead if $n_2\neq 0$, an ``asymmetry" in the left/right reflection coefficient starts to develop 
\cite{RLKCEC11}. We would like to note that the ${\cal PT}$ arrangement considered here is fundamentally different 
from that encountered in distributed feedback lasers (DFBs) \cite{KS71}. In DFB systems both the index and gain/loss 
profile vary in phase and thus no ${\cal PT}$-symmetry breaking is possible. 

At $n_1=n_2$, this asymmetry becomes most pronounced. Even more surprising is the fact that at the Bragg point $\delta=0$, 
the transmission is identically unity i.e. $T=1$, while the reflection for left incident waves is $R_L=0$ (see Fig. 
\ref{fig2}). This is a direct consequence of the ${\cal PT}$- nature of this periodic structure. At the same 
time, the reflection for right incident waves grows with the size $L$ of the sample as
\begin{equation}
\label{Rright}
R_R = L^2\left(k{n_1\over n_0}\right)^2 ({\sin(L\delta )\over L\delta})^2\,\,\,\underrightarrow
{\scriptstyle {\delta\rightarrow 0}} \,\,\,L^2\left(k{n_1\over n_0}\right)^2
\end{equation}
Such quadratic increase of the field intensity is a hallmark of exceptional point dynamics \cite{ZCFK10}. This behavior 
is directly confirmed by our numerical simulations. We will refer to this phenomenon as 
{\it unidirectional reflectivity}. Furthermore, Eqs. (\ref{tcoef}) indicate that a transformation $n_2 \rightarrow 
-n_2$, reverts the reflectivity of the system, allowing for reflectionless behavior
for right incident waves i.e. $R_R=0$, while the reflection from the left $R_L$ is now following the prediction of
Eq. (\ref{Rright}). In other words, the phase lag between real and imaginary refractive index dictates the unidirectional
reflectivity of the system.

\begin{figure}[h]
\includegraphics[width=1.0\columnwidth,keepaspectratio,clip]{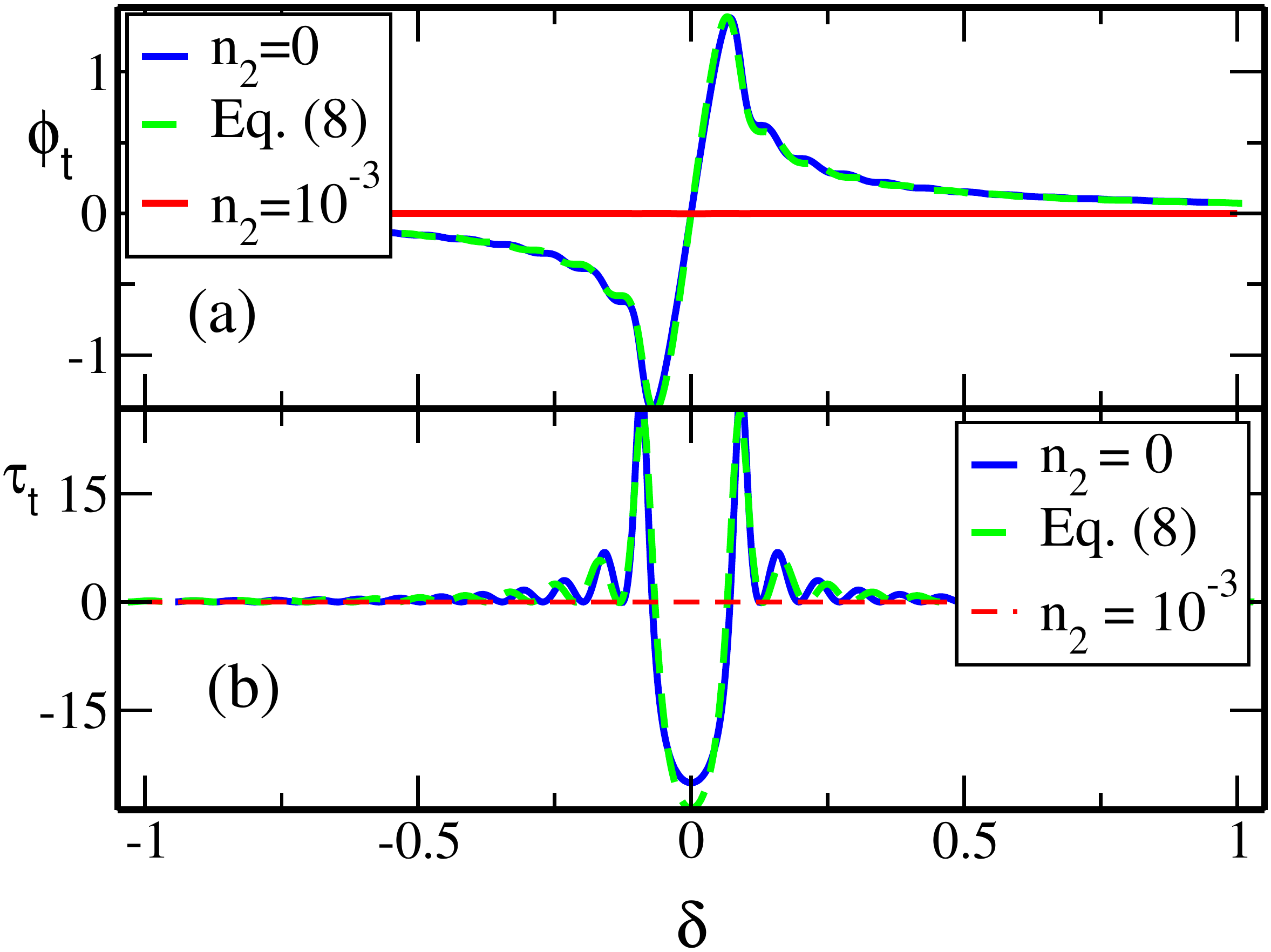}
 \caption{
   \label{fig3}
   (color online) (a) Transmission phase $\phi_t$ as a function of the detuning $\delta$ for the ${\cal PT}$-periodic system of 
Fig. \ref{fig2} \cite{numerics}. For comparison we plot together with the results of the ${\cal PT}$-exceptional 
point ($n_2=n_1$),
also the transmission phase for the passive structure ($n_2=0$). (b) The corresponding transmission delay times $\tau_t$ as a function
of detuning.
   }
\end{figure}

Reflectionless potentials in one-dimensional scattering configurations are not in general invisible. This is due to the fact that 
the phase of the transmitted wave might depend on energy, thus leading to wavepacket distortion after the potential barrier.
In this respect, a transparent potential can be detected from simple time-of-flight measurements. 
It is therefore crucial to examine the phase $\phi_t$ of the transmission amplitude $t=|t|\exp(i\phi_t)$ and compare 
it with the phase acquired by a wave propagating in a grating-free environment ($\phi_t=0$)\cite{phase}
Using Eq. (\ref{cmt}), we deduce that 
the phase $\phi_t$ close to the Bragg point is
\begin{equation}
\label{phase}
\phi_t = \arctan \left(-{\delta \over \lambda} \tan(\lambda L) \right) + L\delta\,.
\end{equation} 
At $n_1=n_2$ we find that $\delta=\lambda$, which results to a transmission phase $\phi_t=0$. Thus interference measurements
will fail to detect this periodic structure. Although the above theoretical analysis is performed close to the Bragg 
point $\delta\approx 0$, our numerical results reported in Fig. \ref{fig3}a, indicate that these effects valid over a
very broad range of frequencies. For comparison, we also report in Fig. \ref{fig3}a, $\phi_t$ for the case of a 
passive ($n_2=0$) Bragg grating.

Next, we analyze the dependence of the transmission delay time $\tau_t\equiv d\phi_t/dk$ \cite{LM94,HFF87}, on the detuning 
$\delta$. This quantity provides valuable information about the time delay (or advancement) experienced by a transmitted wavepacket 
when its average position is compared to the corresponding one in the absence of the scattering medium. Using Eq. (\ref{phase})
we find that at the spontaneous ${\cal PT}$-symmetry breaking point the transmission delay time is $\tau_t=0$. In Fig. 
\ref{fig3}b, we show results for a ${\cal PT}$-structure at $n_1=n_2$ together with those expected from the passive case.

\begin{figure}[h]
\includegraphics[width=1.0\columnwidth,keepaspectratio,clip]{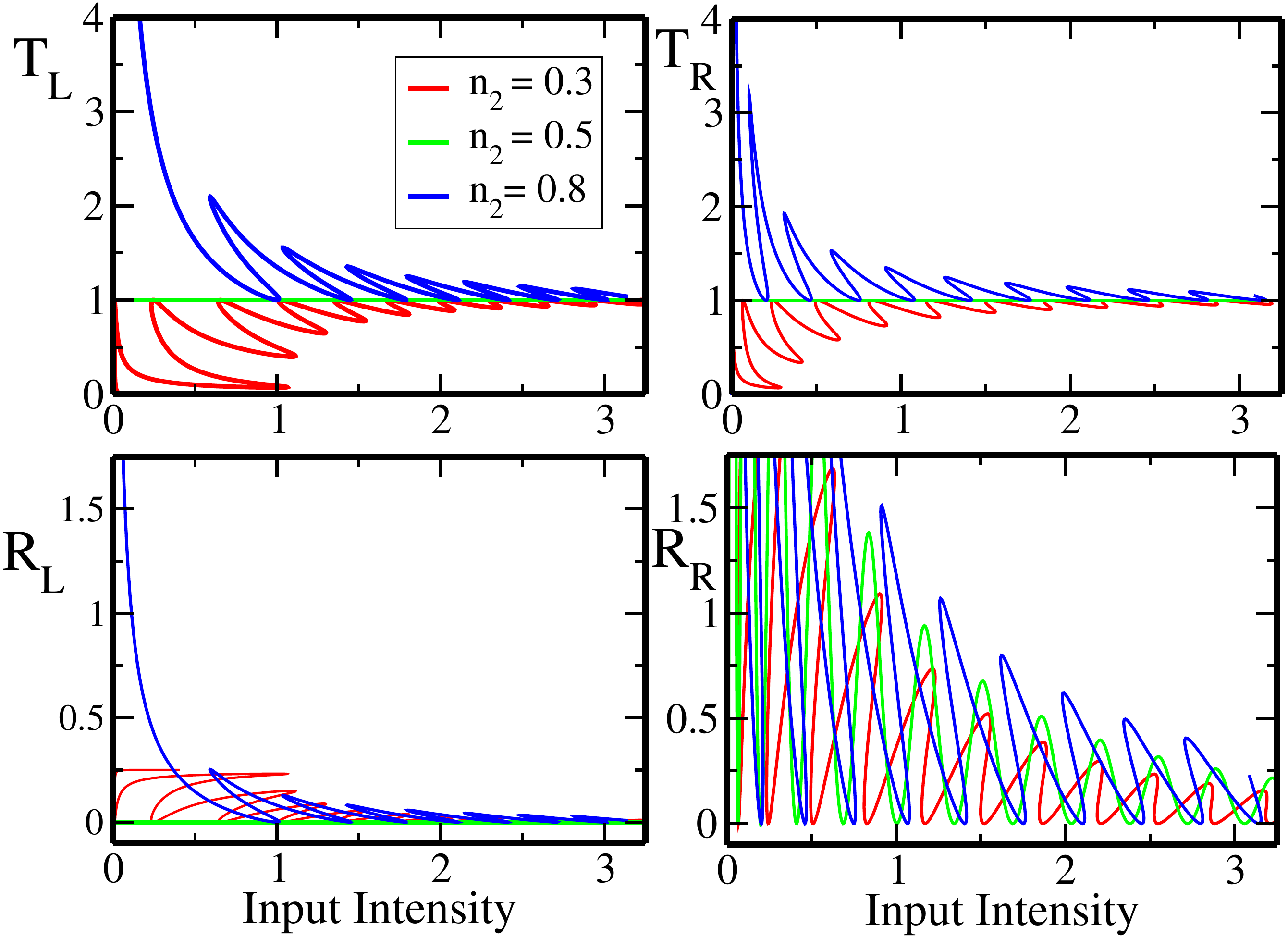}
 \caption{
   \label{fig4}
   (color online) Transmission and reflection coefficients in the nonlinear regime vs. left (left column)
and right (right column) input intensities for $\delta=0$ (the same behaviour is observed for other  values 
of $\delta$). 
The parameters used are $n_0=1$, $n_1=0.5$, and $L=7$. Three different values of $n_2$ (below, above 
and at the exceptional point) are used. If $n_2\neq n_1$ one can observe the standard bistability behavior of non
-linear media. For $n_2=n_1$, the bistability disappears, signifying the appearance of the spontaneous ${\cal PT}$ 
symmetric breaking point. At this point $R_L=0$.
   }
\end{figure}

It is also interesting to investigate the robustness of the above phenomena in the presence 
of Kerr nonlinearities. To this end, we assume the presence of a Kerr term in the refractive index profile i.e. 
$n(z) = n_{0} +n_{1} \cos (2 \beta z)+ i n_{2} \sin (2 \beta z) +\chi |E(z)|^2$. By decomposing the optical field into 
two counter-propagating waves and by considering only synchronous terms \cite{WMG79,RLKCEC11}, we can then
obtain a set of equations describing the field envelopes ${\cal E}_b(z)$ and ${\cal E}_f(z)$, 
in terms of Stokes variables \cite{Stokes} 
\begin{eqnarray}
\label{stokes}
{\dot S_0}(z) = 2 \kappa S_3 \,\, ; \,\, {\dot S_1}(z) = 2 g S_3 \,\, ; \,\, 
{\dot S_2}(z) = 2 \delta S_3 -3 \rho S_0 S_3 \nonumber\\
{\dot S_3}(z) = - 2 \delta S_2 +3 \rho S_0 S_2+2\kappa S_0 - 2g S_1\quad\quad\quad\quad\quad\quad\quad\,\,
\end{eqnarray}
where $\rho=k\chi/n_0$, $\kappa=kn_1/2n_0$ and $g=kn_2/2n_0$. It can be shown \cite{RLKCEC11} that this non-linear 
system has the following conserved quantities $gS_0-\kappa S_1= {\cal C}_1$, $3\rho gS_0^2-4\kappa \delta S_1 + 
4 \kappa g S_2 = {\cal C}_2$. Using these constants of the motion, one can solve exactly Eqs. (\ref{stokes}). Because 
of lack of space we will not discuss the detail derivations here \cite{RLKCEC11} but rather cite the final results 
for the transmission and reflection coefficients
\begin{eqnarray}
T_{L}  =   \frac{(\kappa + g) S_{0}({L\over 2}) - {\cal C}_1} {(\kappa + g) S_{0}(-{L\over 2}) - {\cal C}_1}& ; &
R_{L}  =   \frac{(\kappa - g) S_{0}(-{L\over 2}) + {\cal C}_1} {(\kappa + g) S_{0}(-{L\over 2}) - {\cal C}_1}\nonumber\\
T_{R}  =   \frac{(\kappa - g) S_{0}(-{L\over 2}) + {\cal C}_1} {(\kappa - g) S_{0}({L\over 2}) + {\cal C}_1}& ; &
R_{R}  =   \frac{(\kappa + g) S_{0}({L\over 2}) - {\cal C}_1} {(\kappa - g) S_{0}({L\over 2}) + {\cal C}_1}\nonumber
\end{eqnarray}
In contrast to the linear case, now $T_L\neq T_R$ for $n_1\neq n_2$ indicating a diode action \cite{RLKCEC11,RKGC10} 
(see Fig. \ref{fig4}) However, of interest here is the behavior of the system at the exceptional point $n_1=n_2$. 
We find that $T_L=T_R=1$, while $R_L=0$, as in the linear case. These results are valid for any input intensity as 
shown in Fig. \ref{fig4}. At the same time we have found that the transmission phase is again independent of
the detuning $\delta$ and equal to $\phi_t=0$. We thus conclude that the phenomenon of unidirectional invisibility of 
${\cal PT}$-periodic system at the exceptional point is entirely unaffected by the presence of Kerr non-linearities.

We have shown that the interplay of Bragg scattering and ${\cal PT}$ – symmetry allows for unidirectional invisibility 
which can be observed over a broad range of frequencies around the Bragg point. This process was found to be robust 
against perturbations. In the presence of nonlinearities this unidirectional invisibility still persists and non-
reciprocal transmission is possible. Of interest will be to investigate if these phenomena can also occur in higher 
dimensions and under vectorial conditions.

\begin{acknowledgments}
Useful discussions with V. Kovanis are acknowledged. (ZL), (HR) and (TK) acknowledge support by a 
grant from AFOSR No. FA 9550-10-1-0433 and by the US-Israel Binational Science Foundation, Jerusa\-lem, Israel.
\end{acknowledgments}



\begin{thebibliography}{99}
\bibitem{numerics} We have check (via exact numerical evaluations of $T$, $R_L$, $R_R$ and $\phi_t$, using Eq. 
(\ref{Helmholtz})), 
the generality of our results for other Bragg scattering potentials (e.g. ${\cal PT}$-slab grating)
\bibitem{phase}We do not consider the trivial phase $kL$ associated with free propagation.
\bibitem{Stokes}The Stokes variables are defined as $S_0=|{\cal E}_f|^2+
|{\cal E}_b|^2, S_1=|{\cal E}_f|^2- |{\cal E}_b|^2, S_2={\cal E}_f {\cal E}_b^* + {\cal E}_f^* {\cal E}_b, S_3=i 
({\cal E}_f^* {\cal E}_b - {\cal E}_f {\cal E}_b^*)$ and they satisfy the identity 
$S_1^2+S_2^2+S_3^2=S_0^2$.
\end{thebibliography}
\end{document}